# TICE & normalisation, pour une rénovation universitaire dans les pays du Sud : le cas de l'Afrique francophone subsaharienne

Mokhtar BEN HENDA

**Résumé :**

La rénovation universitaire est un fait récurrent engendré par les innovations technologiques permanentes et les nouveaux modes d'organisation des universités et des offres de formation. Les normes d'interopérabilités technologiques y jouent un rôle déterminant, non pas uniquement en apportant des plus-values sur l'économie d'espace et de temps, mais aussi en changeant les modèles pédagogiques et les processus d'acquisition des connaissances. Les pays d'Afrique subsaharienne présentent un cadre institutionnel et un type d'organisation universitaire particulier qui nécessite une relecture en profondeur de leurs modes opératoires pour une meilleure relance sur la voie de la rénovation par les TIC. Ce document propose des pistes de réflexion et des cadres d'action qui focalisent sur les acquis des normes et des standards d'interopérabilité.

**Mots clés**

Rénovation universitaire, interopérabilité, normes et standards, Afrique subsaharienne, TICE

## 1. Cadre général

Cette contribution prend source dans trois études récentes qui exposent l'état de l'art de l'enseignement supérieur et de la recherche scientifique en Afrique Subsaharienne. La première étude présentée pendant l'atelier sur le financement de l'enseignement supérieur et de la recherche tenu à Cotonou, du 2 au 4 juillet 2007, Borel Foko, analyste des politiques éducatives au Pôle de Dakar de l'UNESCO-BREDA (Bureau régional de l'Unesco pour l'éducation en Afrique), met en évidence 3 défis majeurs à l'enseignement et à la recherche en Afrique subsaharienne :

1. Le besoin de répondre à une demande sociale importante en éducation qui tendra à s'accroître tenant compte de la population africaine la plus jeune du monde ;
2. La nécessité de répondre à des exigences de qualité pour s'aligner sur les standards mondiaux des offres de formation ;

3. L'exigence de répondre aux besoins économiques d'une optique d'expansion de l'enseignement supérieur et de la recherche pour avoir plus de locaux, plus d'enseignants et plus de ressources de meilleure qualité.

Dans la deuxième étude, présentée en 2007 comme un travail de thèse ([1]), Pierre-Jean Loiret élabore une radioscopie de l'état de l'art de l'enseignement à distance en Afrique subsaharienne et avance des statistiques relatives à la pénétration et les usages des TIC dans les milieux africains de l'enseignement et de la recherche. Il pose la question de comment l'enseignement à distance dans le supérieur africain arrivera-t-il à trouver sa voie entre les « dynamiques du dehors » ou internationales (coopérations étrangères et organismes internationaux) et les « dynamiques du dedans » ou nationales (pression démographique et sociale, coûts, innovation). L'étude a ciblé 158 enseignants chercheurs en Afrique de l'Ouest (Bénin, Burkina, Cote d'Ivoire, Guinée, Mali, Niger, Sénégal, Togo).

La troisième étude est publiée par l'ADEA-RESAFAD-UNESCO en 2007 sous forme d'un rapport dirigé par Jacques Guidon et Jacques Wallet sur la FAD dans 14 pays d'Afrique Subsaharienne francophone ([2]). Parmi les constats de cette étude, le niveau de développement des télécommunications en Afrique subsaharienne est le plus faible du monde. En 1990, l'Afrique sub-saharienne comptait 30 établissements d'enseignement supérieur privé, 85 en 1999, plusieurs milliers aujourd'hui auxquels il faudrait ajouter les centaines d'officines nées du « e-Learning » dans un environnement « sans frontière ».

Les auteurs de cette étude sont convaincus que les technologies ne permettent pas de résoudre à elles seules les problèmes structurels dans l'éducation. « *Si les TICE et la FOAD ne sont pas susceptibles de résoudre tous les problèmes systémiques … ils peuvent permettre certains progrès sur des domaines d'intervention et des besoins plus ciblés de formation, et permettra la rénovation de certaines pratiques pédagogiques…* » ([3]).

Ces trois études étaient largement convergentes sur un certain nombre de points caractérisant la réalité de l'Afrique subsaharienne par rapport à l'éducation et les technologies éducatives. Ceci a permis d'établir les quelques constats suivants :

---

[1] Pierre Jean Loiret. EAD en Afrique de l'Ouest. Novembre 2007

[2] J. Wallet. FAD en Afrique Subsaharienne francophone, UNESCO, 2007

[3] J. wallet. Op. Cit.

- La demande sociale pour l'enseignement supérieur en Afrique francophone va plus que doubler d'ici à l'horizon 2015 (facteur 2,6 entre 2004 et 2015). La forte croissance démographique (population jeune) engendre une demande sociale importante pour l'enseignement supérieur. Cette demande va exiger entre autres, un besoin d'enseignants de rang magistral et de plus grandes capacités d'accueil ;

- Les politiques éducatives nationales et au niveau des régions souffrent d'une faible articulation entre elles pour une meilleure harmonisation des systèmes d'enseignement et de recherche spécifiques au contexte subsaharien.

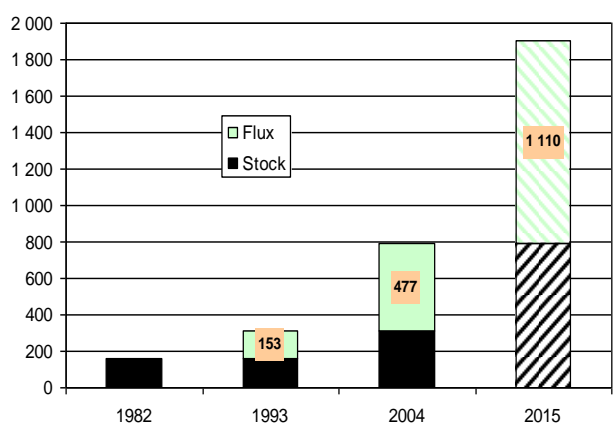

Tableau 1 Accroissement de la demande sociale pour l'enseignement an Afrique
*(Source : Borel Foko, Pôle de Dakar (UNESCO-BREDA). Juillet 2007)*

D'un point de vue infrastructures technologiques et logistiques financières, la situation n'est pas moins critique. Ni les ordinateurs ni les réseaux (particulièrement Internet) n'ont apporté les solutions idoines à la restructuration universitaire, car la qualité et la quantité des accès aux TIC par les enseignants chercheurs africains, comme quantifié par Loiret, annulent l'ancien discours diabolisant l'outil comme source des maux des universités africaines. En effet, Loiret estime la situation actuelle de la pénétration et de l'usage des TIC dans les universités africaines comme suit :

- 81% des enseignants chercheurs déclarent disposer d'un PC,
- 72% affirment disposer d'un ordinateur au sein de leurs institutions
- 27% sont connectés à Internet à domicile

- 70% accèdent à Internet via les Cybercafés
- 32% mettent leurs cours en ligne
- 52% ont suivi des formations EAD
- 52% ont modifié leurs pratiques pédagogiques grâce aux TIC

Les finances n'ont pas résolu, non plus, la rénovation universitaire malgré les fonds drainés par les projets d'aide au développement dans le secteur éducatif, financés par des aides et des crédits de la Banque Mondiale et le Fond Monétaire International. Toutes les formes de conventions, de collaborations, de réseaux, de consortiums régionaux n'ont pas donné lieu à la gouvernance recherchée car ces premiers restent fragmentaires et parcellaires.

Où se situe donc l'intelligence de la réorganisation et de la rénovation universitaire en Afrique subsaharienne ? Est-ce dans les outils et le Transfert de technologies ? Est-ce plutôt dans les moyens financiers et les subventions d'aide au développement de la part des bailleurs de fonds ? Serait-ce plutôt une question de compétences locales et de référentiels métiers ? Peut-on rechercher la solution dans les politiques éducatives et les nouvelles réformes du LMD ([4]) ? Ou est-ce finalement une crise dans les rapports historiques Nord/Sud qu'il faudrait redéfinir sur des assises plus élaborées ? Serait-ce tout ça à la fois ?

## 2. Les TICE dans le cadre d'une rénovation de l'enseignement et de la recherche en Afrique Subsaharienne

Essayer de trouver une réponse aux questionnements précédents est de loin plus complexe qu'une simple énumération de facteurs pouvant servir de levier de développement universitaire et de recherche académique. Nous proposons néanmoins d'étudier un ensemble de facteurs endogènes et exogènes à la réalité africaine pour essayer de donner des pistes de réflexion pour une stratégie d'action lente mais durable. Car à notre sens, il n'y a pas de solution 'clé en main' (*Top/Down*), encore moins une solution 'miracle' qui abrègerait les étapes de la mise en place d'une réelle dynamique de réforme. Il s'agit plutôt d'un processus long et minutieux qui engage le plus grand nombre d'acteurs et touche un large panel de décideurs dans plusieurs domaines de responsabilités.

---

[4] Nicole Loraux, Corinne Sliwka. Formateurs et formation professionnelle : L'évolution des pratiques, Volume 2. Éditions Lamarre, 2006. 355 p. ISBN : 275730030

On se poserait, dès lors, la question que si toutes les pistes décrites par ces études n'ont pas donné lieu à des résultats probants, de quel côté faudrait-il aller encore rechercher des facteurs d'appui à la réorganisation universitaire en Afrique Subsaharienne pour qu'elle affiche des indicateurs plus éloquents que ceux avancés dans ces études ?

Nous essaierons dans ce papier d'apporter un bout de réflexion sur cette question en détaillant une piste qui, à notre sens, pourrait contribuer à repenser même modestement les stratégies en cours pour atteindre des résultats plus consistants. Cette piste focalise sur les solutions technologiques, non pas dans un esprit de transfert de technologies devenu désormais caduque, mais dans une approche plus active qui implique une contribution plus engagée des acteurs universitaires locaux et une stratégie plus intégrative et de collaboration. Il s'agit d'une stratégie de redéploiement des solutions technologiques (TIC et TICE) selon une approche globalisante et interactive construite sur trois paramètres essentiels :

- Se fixer des objectifs clairs de transversalité et d'homogénéité des usages au sein de la même université par un renforcement des structures de l'Intranet et des réseaux locaux. Nous observons jusqu'ici que les expériences de l'informatisation des universités passent souvent par des procédures parcellaires et des étapes préférentielles traitant les différentes activités de façon inégales et disproportionnées. Les TIC ont souvent été introduites pour des besoins de gestion administrative et des tâches de secrétariat configurées via des solutions monopostes. Les aspects pédagogiques et de diffusion des connaissances ont été dans les meilleurs des cas réduits à des sessions de formation courtes et sans suivi. Or, une solution globale devrait prévoir une gestion des flux au sein de l'université et un partage équilibré entre les différents aspects de la vie universitaire, administratifs et organisationnels, certes, mais aussi pédagogiques et diffusion des savoirs dans lesquels les infrastructures des réseaux Intranet auront un rôle essentiel à jouer ;

- Définir une approche d'organisation commune et globale par la rationalisation des dispositifs technologiques installés dans les différentes structures de l'université. Cela sous-tend une modularité souple dans les choix des solutions technologiques à introduire qui tiendrait compte des critères de l'intégration et de la convergence de ces solutions. Il est souvent observé que des solutions d'informatisation sectorielle ne parviennent pas toujours à s'intégrer dans une conception ouverte et élargie. Un système d'information global peut se construire par petits bouts, sauf que ces bouts doivent correspondre à des modules souples facilement incorporé dans la totalité du dispositif à tout moment ;

- Ces objectifs de transversalité et cette approche d'organisation nécessiterait des outils compatibles qui aideraient à atteindre des niveaux de performance avancée de la rénovation universitaire par les TIC. Aujourd'hui, le milieu informatique est largement doté d'outils et d'applications répondant à ces critères de globalité modulaire, d'intégration des services et de convergence des ressources. Les portails institutionnels et les environnements numériques de travail (ENT) se démarquent rapidement comme des solutions très à la mode de la rénovation universitaire par les TIC. Ces solutions nécessitent, toutefois, un socle technologique consistant largement ancré dans des dispositifs qui tournent en réseau distribué. C'est justement à ce niveau que les études précédentes ont détecté le maillon faible e la chaine, celui des réseaux distribués fondés sur les technologies des télécommunications et de la transmission des données.

Notre réflexion partirait sur la voie de cette piste collaborative par les TIC pour exposer les finalités et les avantages du recours aux environnements numériques distribués comme alternative possible pour renflouer les bases d'une politique général de rénovation universitaire dans les pays de l'Afrique subsaharienne. Deux environnements précis font l'unanimité dans les stratégies de l'informatique universitaire et des campus numériques : les portails institutionnels et les environnements numériques de travail (ENT).

## 2.1. Les portails institutionnels

Toute université est composée de structures organisationnelles multiples qui gèrent les aspects financiers, administratifs et de scolarité. Ces structures fournissent des services qui ne sont pas forcément coordonnées entre eux. Ceci pose a priori des problèmes de compréhension et de communication entre les acteurs dans chaque service. Ceci poserait aussi des difficultés à chaque opération d'échange de données pour des situations de prise de décision. Parmi les conséquences immédiates à cette différence dans les approches et les sensibilités quant à l'information possédée et distribuée, c'est la duplication des ressources et donc tout le mal à les synchroniser.

La solution que plusieurs universités du Nord ont déjà entreprise, est celle d'une fédération des services et des ressources autours d'un portail institutionnel unique qui offre l'opportunité d'élargir les accès au système d'information institutionnel à un plus grand nombre d'utilisateurs selon des moyens et des mécanismes homogènes en utilisant une interface commune, personnalisée et sécurisée. Un portail institutionnel est un environnement où les informations générales sur chaque structure de l'institution sont disposées selon un format homogène préalablement établi.

Ainsi défini, l'ensemble des processus et des solutions qui seront mis en place par toute université seraient bâties sur un socle unique et transversal. Ce socle serait composé de « briques » techniques et servirait particulièrement pour la gestion des flux des données entre les services (*workflow*), le contrôle des accès distants grâce à un annuaire LDAP centralisé, la gestion électronique de documents, bref la coordination et la synchronisation de toutes activités qui touchent chacun des acteurs inscrit dans le système d'information de l'université. C'est l'un des critères essentiels de l'intégration, de la convergence et de l'interopérabilité souhaitée à travers ce genre de dispositifs ([5]).

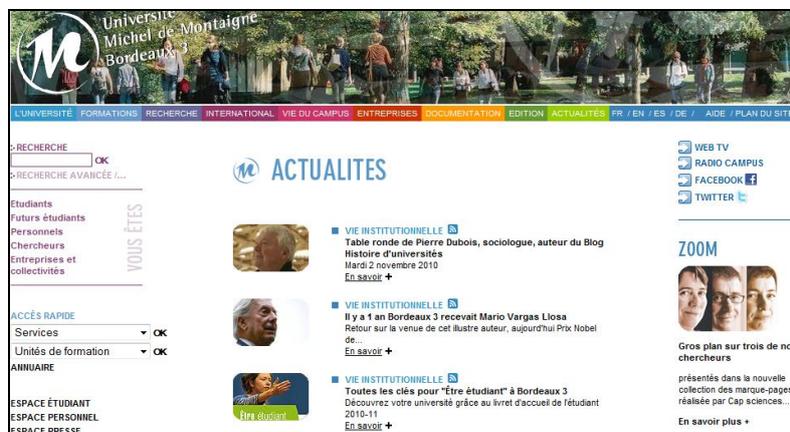
Figure 1: L'exemple du portail institutionnel de l'Université de Bordeaux 3

La réussite de ce socle dépendra de beaucoup de facteurs clés. Le plus important est sans doute le degré d'harmonisation et d'adaptation de l'ensemble des services de l'université aux caractéristiques fédératrices du système d'information envisagé. Cette caractéristique est essentielle pour fournir aux usagers un ensemble de services autonomes et décentralisés.

---

[5] Denis Berthier. Le savoir et l'ordinateur. L'Harmattan, 2002. 457 p. ISBN : 2747533506

Elle garantira aussi une économie d'échelle puisqu'elle permettrait des gains de temps et d'espace dans l'accès et la conservation des ressources documentaires de l'université.

La réussite d'un portail institutionnel repose aussi sur sa capacité à créer des liens avec d'autres sites et d'autres portails sur les réseaux. Ceci rejoint le principe de la mutualisation des moyens et des ressources dans le cadre des consortiums universitaires, des partenariats ou des stratégies de fusion. Sinon, c'est une des formes de syndication des ressources au profil d'une communauté de pratique présumée recevoir des services externes à travers le portail de l'institution de rattache.

Les portails institutionnels sont généralement équipés d'environnements plus restreints, ceux des environnements numériques de travail (ENT)

### 2.2. Les ENT : un cadre d'intégration des services par les TICE

Les ENT sont aussi des portails accessibles au sein de l'institution ou depuis le domicile à travers un point d'entrée unique contrôle par un identifiant et des droits d'accès. Chaque utilisateur inscrit dans un annuaire institutionnel accède à un espace personnalisé dans lequel il peut exploiter les ressources et les services numériques mise à la disposition de la communauté, chacun selon son statut et ses activités. Un étudiant accèderait à un espace virtuel souvent dénommé « bureau virtuel » dans lequel il gère son emploi de temps, consulte des ressources documentaires et travailler en collaboration avec des membres de son groupe.
Les ENT sont aussi des lieux de rencontre, de communication et de collaboration (synchrone ou asynchrone) entre les personnels administratifs, les étudiants et les enseignants. Ils jouent ainsi un double rôle :

- Un rôle essentiellement pédagogique avant tout, puisqu'il va faciliter aux enseignants et aux étudiants l'accès à des ressources scientifiques et à des bases de données documentaires. Il servira aussi de cadre d'apprentissage grâce auquel l'enseignant dépose se ressources en accès permanent, effectue les épreuves de contrôle et dépose les corrigés d'exercices et des travaux ;

- Un rôle de cadre de communication entre étudiants, enseignants et administration pour gérer et exploiter les outils de communication intégrés (portfolios, chat, messagerie, forum, wiki, agenda, annuaire etc.).

Dans cette perspective, il est indispensable que l'institution accomplisse une action de mise à niveau des réseaux informatiques à l'intérieur des établissements et étudient les possibilités de connexion depuis l'extérieur. Il en va de la réussite de l'implantation de l'ENT et surtout de son efficacité.

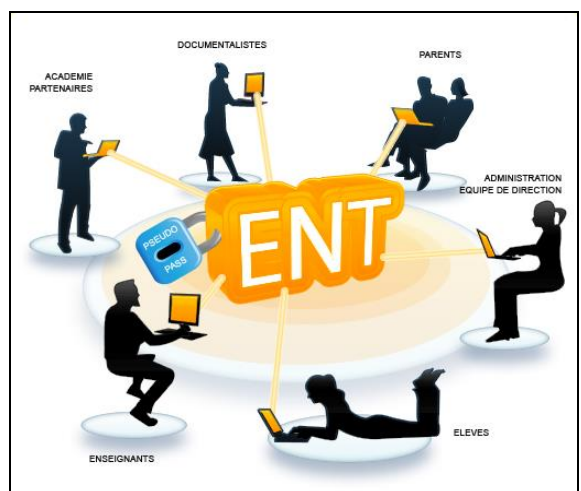
Figure 2 : Configuration d'un ENT
*(Source : Documentation Dokeos, http://www.dokeos.com/fr/node/687)*

Ces quelques mesures sont en réalité la partie visible d'une chaine plus complexe de préparatifs institutionnels, techniques, financiers, organisationnels et même juridiques. Tant pour les institutions universitaires subsahariennes que pour toutes autres institutions dans n'importe quel pays du monde, ces mesures restent sans aucun doute tributaires d'une conjoncture générale et d'un cadre réglementaire plus large desquels aucune institution universitaire ne peut se soustraire. Il s'agit entre autres d'une politique nationale pour l'enseignement supérieur et la recherche, d'une stratégie nationale pour les technologies de l'information et de la communication et l'informatisation des structures universitaires, d'une conjoncture de télécommunications et de l'infrastructure d'accès aux réseaux internationaux, particulièrement le réseau Internet, etc.

Or, si l'on reste tributaire de tous ces facteurs à la fois, il est certain que l'on ne parvienne jamais à disposer d'un cadre optimum dans lequel tous les facteurs sont favorables à l'installation d'un ENT. Une politique de rénovation universitaire est un processus long et complexe. Sa maîtrise nécessite une stratégie d'étape qui sait anticiper les changements et prédire les innovations structurelles et technologiques. Elle devrait alors focaliser sur les phases réalisables selon des agendas réalistes sans perdre de vue des facteurs clés qui rendent toute modification, mise à jour ou réforme ultérieure facilement intégrable et modulable par rapport à tout le dispositif. Ces facteurs sont impérativement ceux de l'interopérabilité, l'intégration et la convergence. Ces caractéristiques qui sont désormais au cœur même de tout dispositif technologique moderne sont à leurs tours tributaires d'une action transversale, celle de la standardisation et de la normalisation des solutions technologiques, des ressources numériques et des services à valeurs ajoutées sur les réseaux.

### 3. L'interopérabilité au cœur des processus de rénovation universitaire

En effet, pour que toute intégration de services et de ressources fonctionne, il est impératif qu'elle réponde à des exigences d'interopérabilité par les normes et les standards technologiques. Pour que l'intégration fonctionne, il faut un minimum d'homogénéité, de compatibilité, bref de standardisation et de normalisation. Sauf qu'il faut bien se poser la question sur l'objet de cette normalisation ou de cette standardisation. Á vrai dire, ce genre de question a été largement posé dans d'autres secteurs d'activités qui se sont rapidement appropriés des innovations technologiques comme le secteur économique, le domaine de la santé, les administrations publiques et tant d'autres qui disposent aujourd'hui de leurs propres systèmes de réseaux internes et externes connectés sur les réseaux des télécommunications mondiaux.

L'éducation est l'un des secteurs qui affiche encore une résistance à ce renouveau technologique par les normes. Cette réactivité relative est toutefois compréhensible si l'on considère que le domaine de l'éducation et des technologies éducatives est un croisement de disciplines qui apportent chacune son lot de spécificité et de contraintes. Mais, l'innovation est déjà en marche véhiculant dans ses sillons des pans entiers de normes et de standards désormais très nécessaires pour garantir les acteurs de l'interopérabilité, de l'intégration et de la convergence.

En effet, les TIC et les réseaux sont largement tributaires de normes et de standards qui garantissent des niveaux d'interopérabilité maximale dictés par les préoccupations industrielles et les valeurs marchandes des produits. Des standards de fait caractérisent désormais tous les produits technologiques du marché (e.g. Microsoft). Des normes de consensus sont également adoptées par des structures du domaine public comme le réseau Internet (e.g. W3C, IETF). Des instances de normalisation internationales (e.g. ISO, IUT) ont aussi apporté leurs lots de normes pour cadrer les infrastructures technologiques et des télécommunications dans le monde. Le contexte universitaire a fini par céder à l'idée de prévoir ses propres normes et standards technologiques qui lui apporteraient les mêmes valeurs ajoutées observées dans d'autres secteurs d'activité.

Les avantages de l'interopérabilité par les normes sont désormais palpables à plus d'un niveau dans la dynamique opérationnelle de toute institution universitaire. Elles peuvent concerner le mode de fonctionnement de l'institution, ses offres de formation, la configuration de es ressources informatiques, et plus important que tout, ses processus pédagogiques. Pour chacun de ces aspects de l'interopérabilité d'un dispositif technologique universitaire, des référentiels normatifs internationaux sont déjà définis ou en phase de construction.

### 3.1. L'interopérabilité organisationnelle

Le recours aux normes et standards technologiques pour l'organisation des modes opératoires des universités passe obligatoirement par des référentiels généraux de l'informatique et des télécommunications. S'agissant essentiellement des technologies sur lequel se construisent les services numériques à mettre en synergie, un besoin de repenser le déploiement des TIC dans les universités est essentiel en fonction des besoins génériques d'intégration, d'homogénéité et de transparence sur différentes couches : gestion, scolarité, pédagogie, documentation, maintenance, communication, etc. Sur ce point, les ENT sont une fois de plus une alternative intéressante s'ils reproduisent les caractéristiques d'un travail en réseaux, avec les logiques de partage et de collaboration héritées des applications *groupware* et des systèmes de *workflow*. L'interopérabilité dans ce cas précis permettraient aux différents services de d'échanger des ressources complémentaires sur la vie de l'université sans être obligé d'utiliser les mêmes applications ni les mêmes formats de données. Les services garderont leurs autonomies fonctionnelles mais pourront profiter de l'interopérabilité de leurs dispositifs technologiques pour communiquer, échanger et mutualiser des ressources. Il y a là un facteur d'économie d'échelle important puisqu'elle permettrait d'éviter les doublons des ressources et les doubles efforts de traitement des données.

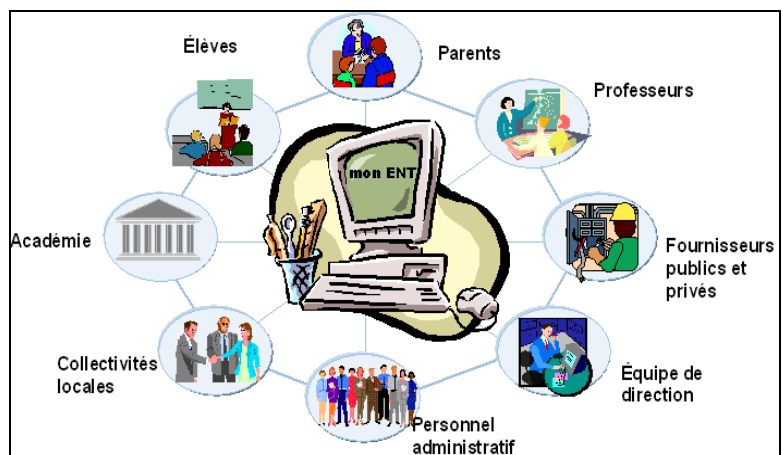

Figure 3 : une interopérabilité organisationnelle universitaire organisée autour d'un ENT
*(Source : http://college-clostardif.etab.ac-caen.fr/sites/college-clostardif.etab.ac-caen.fr/IMG/png/ent2.png)*

### 3.2. L'interopérabilité des offres de formation

Les offres de formation sont l'essence même de l'identité des institutions universitaires. Elles déterminent l'envergure de l'institution, la nature de ses cursus, la typologie de ses diplômes et la qualité de ses formations. Les institutions universitaires, surtout celles qui entrent dans des processus d'autonomie, se donnent de plus en plus à soigner cet aspect identitaire pour plusieurs raisons. C'est un facteur de compétition pour drainer un public étudiant de meilleures qualités. C'est aussi un moyen de développer des partenariats pour faire face aux coûts des ressources pédagogiques. Les universités se mettent de plus en plus en consortium pour partager des ressources documentaires scientifiques comme les abonnements groupés à des revues scientifiques. Les offres de formation sont aussi un miroir auprès du marché de l'emploi et des opportunités de travail pour les diplômés.

En définitive, les universités ont tout l'intérêt à harmoniser leurs politiques éducatives avec un minimum de consensus dans le « comment » et le « quoi » faire. Cela nécessite une maîtrise de son propre environnement institutionnel pour identifier l'existant, définir les besoins et proposer les solutions. Une offre de formation doit être consignée dans un descriptif cohérent qui constitue le référentiel de l'institution élaboré selon des standards internationaux comme le CDM (*Course Description Metadata*) (⁶).

Le CDM est un modèle européen de description des offres de formation. Il se décline en un descriptif qui obéit à une structuration réglementée qui fournit des informations sur les établissements, le personnel enseignant et les programmes de formation.

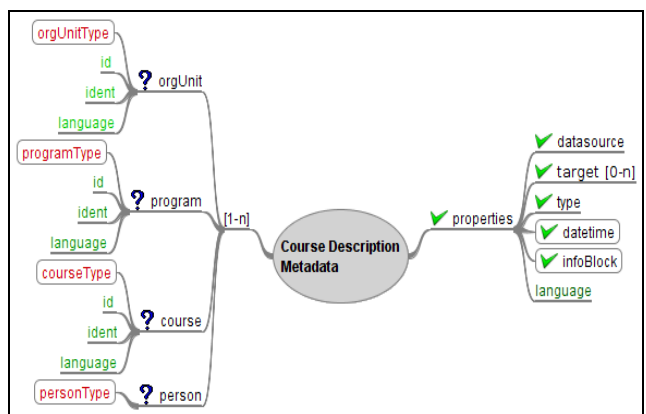

Figure 4 : les composantes clés d'un schéma CDM

Le schéma CDM est structuré en 4 entités descriptives :

1) Une entité organisationnelle (**orgUnitType**) qui concerne l'organisation responsable du déroulement des cours (les universités, les facultés ou les établissements scolaires). Elle se décompose en 19 éléments de description d'un niveau inférieur ;

2) Une entité de description des programmes d'études (**programType**) qui couvre l'ensemble des cours ou de modules conduisant à un diplôme (e.g. LMD). Cette entité est détaillée en 24 éléments de description de niveau inférieur ;

3) Une entité de description des enseignements (**courseType**) qui composent un programme donné. Elle est composée d'un ou de plusieurs cours avec un emploi de temps, des activités d'enseignement et un ou plusieurs épreuves ou examens. Elle contient 27 éléments de description de niveau inférieur ;

---
⁶ Rory McGreal. *Online education using learning objects*. Routledge, 361 p. ISBN : 20040415335124

4) Une entité pour décrire les ressources humaines (**personType**). Il s'agit des coordonnées des personnes impliquées dans l'organisation et le déroulement des composantes susmentionnées. Elle contient 6 éléments d description de niveau inférieur.

Les offres de formation sont aussi rendues utiles ans une perspective de communication, car elles permettent de renseigner les étudiants, les enseignants et le personnel administratif sur des entités qui ne sont pas de leurs ressorts directs. Cette information permettrait aux étudiants de faire des comparaisons de cursus et de nature de diplômes à condition, bien évidemment, que toutes les offres de formation soient signalées, et qu'elles le soient de manière homogène et uniforme. Le rôle des normes dans ce cas est d'assurer cette uniformité descriptive des offres de formation pour les rendre accessibles aux publics concernés.

### 3.3. L'interopérabilité technologique

Le domaine technologique est le plus concerné par les normes d'interopérabilité. Toutes les technologies que nous utilisons, tant dans le domaine public que dans des secteurs spécialisés, sont soumises à des règles normatives de tout genre (spécifications, standards de fait, standards de consensus, normes). C'est un héritage de l'informatique et des télécommunications qui ont su progresser sur ce terrain depuis le siècle dernier. On n'aurait pas des micro-ordinateurs si ses différents périphériques ne respectaient pas des seuils élevés de normalisation pour pouvoir fonctionner dans une construction modulaire des PC. On n'aurait pas un réseau Internet si se protocoles de fonctionnement, d'administration et de contrôle ne sont pas conçus selon des spécifications normatives communes pour permettre à tant de systèmes (outils, logiciels, données) de fonctionner ensemble de manière transparente.

Les ENT ne dérogent pas à cette règle en tant que dispositif technologique composite. En effet, les ENT regorgent de standards et de normes pour le montage et le démontage des composantes du système sur une base d'interopérabilité systématique. Chacun de modules constituant l'ENT, comme la messagerie électronique, la gestion documentaire, les navigateurs Web, les moteurs de recherche, les annuaires LDAP, le portfolio etc., est le résultat d'un ensemble de normes préétablies qui configure les modes de sa portabilité, de son interopérabilité et de sa réutilisabilité.

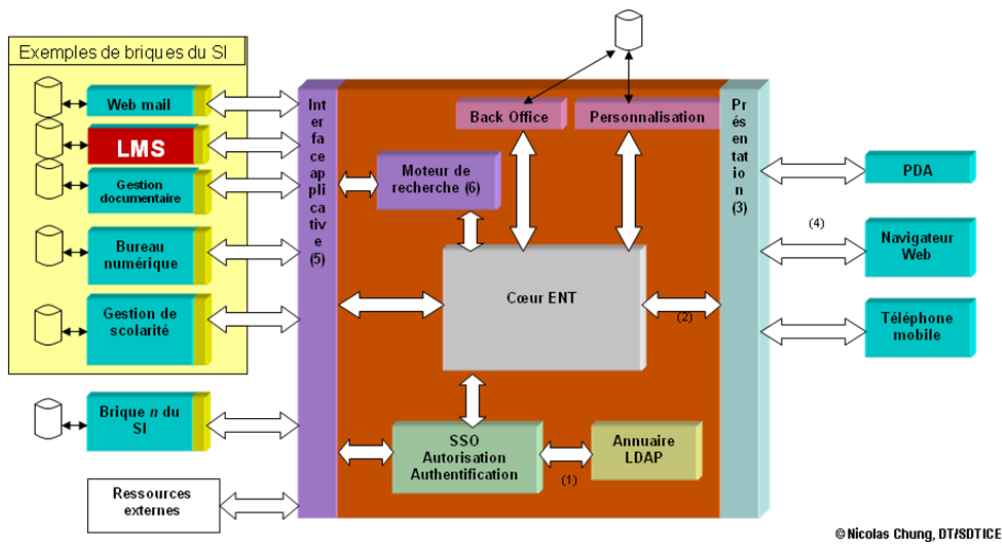

Figure 5 : Un exemple de composantes d'un ENT

### 3.4. L'interopérabilité pédagogique

Vu du côté des contenus et des usages, plutôt que de la quincaillerie et des applications, les ENT ont pu créer des univers d'échange et de communication transparents entre les différents services et acteurs de l'université. On parle de plus en plus des TICE (Technologies de l'Information et de la Communication pour l'Éducation) pour désigner les infrastructures technologiques qui servent à faciliter les activités pédagogiques et de gestion des parcours d'apprentissage. Les TICE proposent aujourd'hui des solutions pour l'organisation des contenus pédagogiques en ligne et la définition des scénarios pédagogiques d'apprentissage.

Or, une part d'harmonisation au sens sémantique et pédagogique reste sans doute à renforcer par les normes et les standards d'interopérabilité. C'est d'ailleurs le grand chantier des normes et des standards des TICE et de la FOAD qui se joue en ce moment à échelle internationale dans un objectif clé, celui de rendre interopérables les systèmes éducatifs et de développer des réseaux sémantiques pour les alimenter en ressources de connaissance et de savoir. La nouveauté de ce créneau et sa charge intellectuelle et pédagogique forte, en font un terrain où les pays du Sud peuvent encore agir pour marquer cette mouvance internationale par leurs propres empreintes culturelles et linguistiques.

Deux aspects de l'interopérabilité pédagogique sont en avance. Il s'agit des ressources de contenus et des scénarios d'apprentissage.

### 3.4.1. Les contenus d'apprentissage

Les contenus sont la matière première de toute action de formation, d'enseignement et d'apprentissage. C'est la raison pour laquelle une attention particulière a été attribuée à ce composant stratégique des dispositifs d'enseignement. Les contenus ont aussi l'avantage de partager des caractéristiques avec les ressources d'information scientifiques et technique. La distinction majeure qui les sépare est le contexte d'usage. Alors que les documents scientifiques et techniques servent de matière première pour une communauté de chercheurs, et donc ne sont pas soumises à des conditions et à un contexte spécifique d'usage, les contenus pédagogiques sont plutôt utilisés dans des conditions d'usage plus cadrés : (les cursus et les épreuves de contrôle de connaissance) et des contextes confinés (séances de cours et plates-formes).

Cette distinction allait jouer sur la nature de description de ces ressources pour des besoins d'usage. En général, parmi les critères qu'une norme de description de contenu est censée garantir, sont les cirières de réutilisabilité, d'adaptabilité, d'interopérabilité et de durabilité. Ces critères s'appliquent aussi aux ressources de contenus pédagogiques, mais celles-ci disposent de critères spécifiques qui préservent leurs spécificités particulières. Dans les schémas de description par métadonnées normalisées, comme le LOM (*Learning Object Metadata*) ou le schéma « *Dublin Core Education* » il y a un consensus pour réserver une branche du schéma de métadonnées à des critères de description propre au contexte de l'éducation. Des critères comme le niveau de difficulté de la ressource, sa densité sémantique, l'état de sa finition, son type, le niveau de scolarité ciblé, etc. sont des paramètres qui ne concernent que le contexte d'apprentissage.

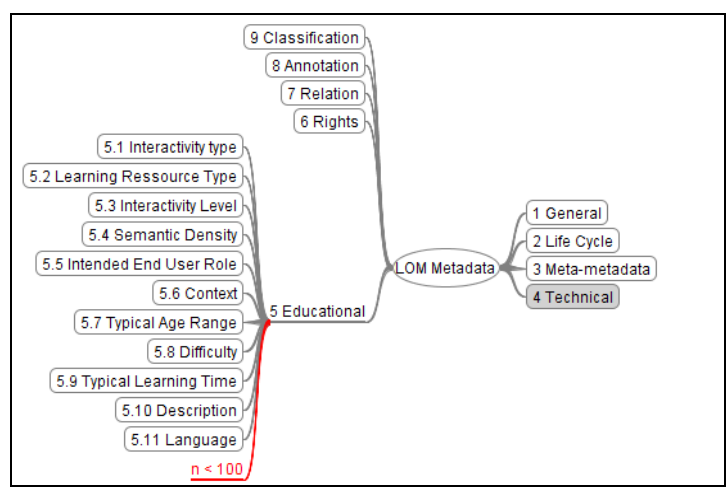

Figure 6 : La branche « *Éducation* » dans un schéma LOM de métadonnées pédagogiques

Ces détails de description des ressources de contenu, augmentent les opportunités de leurs échanges, mutualisation et réutilisation si elles sont conformes à des spécifications, des standards et des normes reconnus et partagés par tous.

### 3.4.2. La scénarisation Pédagogique (LD)

Il faut admettre d'emblée que la scénarisation pédagogique est un domaine où les normes n'ont pas pu avoir un impact aussi remarqué qu'avec les contenus pédagogiques. La scénarisation est au fait un travail de planification de l'activité d'apprentissage, de conception de contenu et d'organisation des ressources. Il existe aujourd'hui des réservoirs de scénarios pédagogiques accessibles sur Internet, mais leur usage est resté limité pour plusieurs raisons. D'une part les enseignants connaissent peu leur utilité. D'autre part, et c'est là l'une des caractéristiques des enseignants et des pédagogues, il y a un rejet à cette forme d'enseigner. France Henri pose dans ce sens une question délicate quand il se demande : « *Peut-on normaliser la conception d'un scénario pédagogique qui suppose une originalité d'action, un souci particulier d'implication de l'apprenant afin qu'il développe une démarche active et des attitudes autonomes* ? » ([7]). Une telle question invite à se demander quelles difficultés freinent le développement des scénarios pédagogiques. Á quel niveau situer la résistance des enseignants à développer des pratiques pédagogiques différentes ?

Á vrai dire, historiquement, la scénarisation pédagogique vient du monde de l'audiovisuel qui a joué un rôle important dans les programmes de formation et d'éducation. La scénarisation médiatique de l'audiovisuel a donné lieu à une nouvelle forme de présentation des savoirs en utilisant les ressources audiovisuelles comme l'image et la vidéo. Cette innovation a poussé vers une reconsidération de l'action d'apprendre et la fonction d'enseigner. La nouvelle conception de l'enseignement par les TICE a réduit la tâche de l'enseignant à un simple régulateur d'interactions entre l'apprenant et les médias, qui sont les porteurs réels du savoir et de la connaissance. La résistance est peut-être à ce niveau. Or, si l'on regarde de plus près les avantages de cette transformation, le fait d'intégrer des technologies de personnalisation de parcours dans les modèles d'enseignement, pourrait être potentiellement utile pour formuler des objectifs pédagogiques et des plans de cours spécifiques selon les besoins des apprenants. À l'heure actuelle, les systèmes ne fournissent pas encore les outils pour identifier les tendances dans les pratiques de conception des apprentissages. Ils n'offrent pas non plus des possibilités pour les enseignants afin de personnaliser les parcours d'apprentissage et de collaborer avec leurs pairs dans le développement de modèles plus efficaces.

---

[7] France Henri, Carmen Compte, Bernadette Charlier. La scénarisation pédagogique dans tous ses débats. Revue internationale des technologies en pédagogie universitaire, 2007. 4(2)

Il est donc important d'offrir aux enseignants des services intelligents, tels que la personnalisation et l'adaptation des contenus, l'appariement des caractéristiques des apprenants avec les modèles d'apprentissage spécifiques, et d'exploiter les informations restituées de la part des enseignants sur l'utilité et la pertinence des objets d'apprentissage ou des modèles de conception pour certains scénarios d'apprentissage.

Plusieurs spécifications et standards de scénarisation des contenus et des parcours sont aujourd'hui appliqués. Les plus connus sont certainement ceux d'IMS (*Instructional management System*) à travers son modèle IMS-LD (*Learning Design*) ([8]).

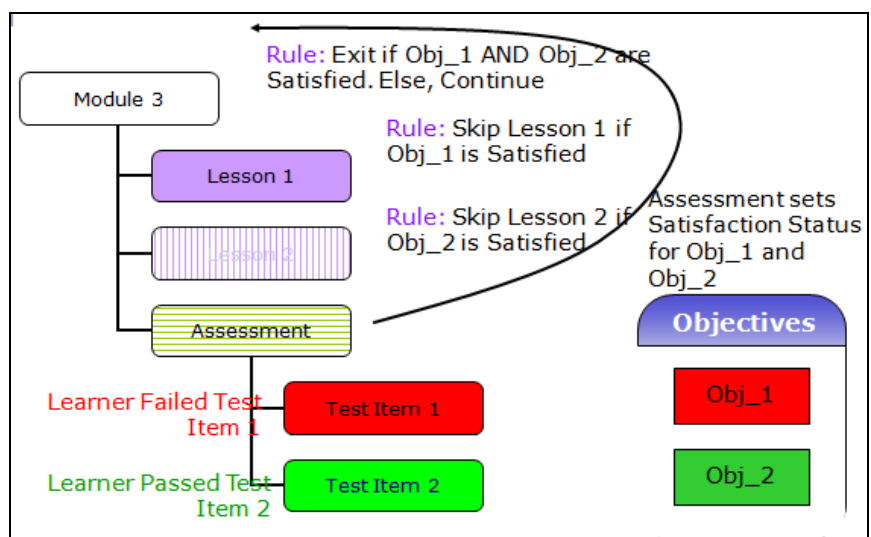

Figure 7 : Un exemple de scénario pédagogique *(Source : IMS)*

L'application des normes des scénarios pédagogiques est, à vrai dire, une étape qui vient au bout d'une chaine antérieure d'actions d'assainissement de l'environnement structurel, technologique,

---

[8] Rob Koper, Colin Tattersall (Dir). Learning Design: *A Handbook On Modelling And Delivering Networked Education And Training*. Springer-Verlag Berlin and Heidelberg GmbH & Co. K., 2005. 412 pages

organisationnel et pédagogique. Il n'en reste pas moins, que la normalisation d'une scénarisation pédagogique, pourrait contribuer à améliorer la rentabilité des dispositifs pédagogiques et promouvoir les actions de collaboration et de partage de ressources entre les institutions universitaires qui y adhèrent. Même situé en aval d'un long processus de rénovation universitaire la scénarisation pédagogique peut être compte parmi les indicateurs de valorisation des normes d'interopérabilité pédagogique dont les intérêts ne sont plus à nier.

**4. Impacts et intérêts des standards pédagogiques**

Les normes sont à la base de l'interopérabilité définie par l'IEEE ([9]) comme « *la capacité de deux ou plusieurs systèmes ou composants à échanger des informations et à utiliser l'information qui a été échangée* ». Elles fournissent un environnement plus prévisible, plus sûr, efficace et fiable qu'un environnement composé de façon aléatoire.

En termes de rapport coût-efficacité, les organisations bénéficient du partage des ressources rendu possible par l'utilisation de normes. La duplication des efforts est réduite et les coûts de développement des ressources de haute qualité peuvent être répartis sur un plus grand nombre d'utilisateurs. Les normes stimulent ainsi l'innovation et le développement de produits et services, tout en réduisant la duplication des efforts et des coûts inutiles.

Les normes procurent aussi un accès facile aux ressources d'information dont elles facilitent le partage et la réutilisation. Elles permettent d'élaborer des approches communes pour trouver, adapter et réutiliser les matériaux de formation et les déplacer entre les différentes collections et plates-formes. Les systèmes d'information basés sur les normes d'interopérabilité accroissent l'efficacité des systèmes de gestion et d'administration de l'information en veillant à ce que les données soient saisies et conservés dans un seul endroit mais utilisées à partir de lieux internes et externes pour des affinités multiples.

Les normes soutiennent l'accès universel à la formation en veillant à ce que les services en ligne puissent être consulté sans trop de problèmes engendrés par l'emplacement physique, les plates-formes logicielles, l'accès au réseau ou de l'invalidité de l'enseignant ou l'apprenant. Le fait d'appuyer l'utilisation des normes permet aux processus de collaboration de se produire indépendamment des frontières organisationnelles, des plates-formes techniques et de l'emplacement géographique. Les apprenants bénéficient aussi de l'amélioration de la qualité des ressources et des possibilités de personnalisation de leurs parcours d'apprentissage.

---

[9] IEEE : *Institute of Electrical and Electronics Engineers*, offre une vaste gamme de publications de qualité et de normes qui rendent l'échange de connaissances techniques et d'information possible entre les professionnels de la technologie.

Ceci dit, les avantages de la normalisation pour les universités subsahariennes ne sont pas de l'ordre du virtuel. Les normes d'interopérabilité sont souvent approchées avec des niveaux d'abstraction qui empêchent leur appropriation par les universités du Sud. Elles sont considérées comme une affaire du ressort des pays développés. Or, une analyse objective des effets escomptés de l'application des normes démontrerait que les pays du Sud sont plus enclins à en tirer des avantages. Il est pourtant vrai qu'une phase de normalisation du contexte universitaire nécessite un nivellement de large échelle pour ramener plusieurs éléments actifs dans la vie d'une institution universitaire à un niveau rapproché de performance pour appliquer dessus des actions de normalisation qui enrichissent leurs capacités d'interopérabilité, d'adaptabilité et d réutilisabilité.

## 5. Pour conclure

Il est donc impératif que les structures universitaires subsahariennes réagissent d'une façon à optimiser leurs ressources et à mieux canaliser leurs efforts vers des logiques de partenariats et de consortiums. Plusieurs pistes sont à prévoir dans ce sens.

Un premier niveau d'action serait celui de la production des ressources pédagogiques. Pour assurer des économies dans les coûts de production des contenus, les normes d'interopérabilité seraient d'un apport considérable dans la construction de réservoirs d'objets pédagogiques interopérables sur le modèle des archives ouvertes et du protocole d'indexation OAI-PMH (*Open Archives Initiative-Protocol for Metadata Harvesting*) ([10]). Les référentiels dans ce domaine existent et il ne s'agit que d'une volonté « politique » et d'une organisation efficace pour entamer un processus de construction de ces réservoirs.

Une autre forme d'économie peut être obtenue en renforçant les capacités d'accueil des institutions universitaires et faire face à la demande sociale accrue d'enseignement, évoquée au début de ce document. La solution serait indéniablement le développement des environnements de formation à distance qui répondraient au mieux aux questions des limites dans les capacités d'accueils et les classes débordantes d'étudiants. Les normes seraient propices pour ce genre de dispositifs technologiques en y introduisant des fortes doses d'interopérabilité organisationnelles et technologiques. Les référentiels de partage de ressources et de gestion des parcours d'apprentissage existent depuis longtemps.

---

[10] Christine Aubry, Joanna Janik. Les archives ouvertes : enjeux et pratiques : guide à l'usage des professionnels de l'information. Association des professionnels de l'information et de la documentation (France) ADBS, 2005. 332 pages

Le déploiement des solutions (intégrales ou partielles) de formation à distance devrait se faire dans le cadre d'une conception concertée au sein d'un consortium régional voir international pour une harmonisation des offres de formation. Cette approche permettrait une mobilité interuniversitaire des étudiants et des parcours individualisés grâce à des offres de formation compatibles. L'avènement du modèle LMD et des systèmes de crédits est un facteur d'appui à cette démarche.

Les conséquences de cette ouverture dans l'approche de rénovation dépasseraient la simple économie d'espace et de moyens. Elle aboutirait à une reconnaissance régionale et internationale des diplômes si elle répondait aux conditions suivantes :

- la définition de référentiels des compétences reconnus et de référentiels métiers validés pour générer des formations professionnalisantes, développer des partenariats avec le secteur privé et répondre aux besoins du marché de l'emploi afin de contribuer concrètement au développement socioéconomique du pays ;
- la recherche des cadres de co-diplômations régionales et internationales pour renforcer les cursus et profiter des échanges d'expertises et de la mobilité des enseignants dans les deux sens ;
- la formation de ressources humaines de haut niveau pour entamer des projets d'envergure et des réformes de grandes ambitions ;
- la collecte des fonds financiers nécessaires pour pérenniser le processus de rénovation ;
- la révision d'une politique d'enseignement générale et de mise à niveau institutionnelle. Il s'agirait de repenser l'architecture des systèmes universitaires (mise en consortiums) avec des structures de réalisation et de suivi.

En définitive, une réorganisation universitaire via les TIC tiendrait compte :

- d'une stratégie rationnelle locale de nivellement par le bas (*Bottom-Up*) pour mettre en synergie les ilots de compétences et des acquis technologiques déjà réalisés et en produire d'autres dans une politique de systémisation globale des procédures de fonctionnement et de maintenance du système universitaire ;
- du développement de consortiums régionaux dans des modes opératoires pratiques (socle/réseaux régionaux) plus que protocolaires ;
- de remonter au créneau international par l'adhésion aux conventions, normes et standards pour garantir une interopérabilité tout en préservant une identité propre.

…/…